\documentclass[9pt,conference]{IEEEtran}

\usepackage[preprint]{waspaa25}

\usepackage{bm} 
\usepackage{amsmath,cite,url}
\usepackage{graphicx}
\usepackage{color}
\usepackage{tipa} 
\usepackage{booktabs}
\usepackage{multirow}
\usepackage[dvipsnames]{xcolor} 
\usepackage{soul, color}
\definecolor{xpink}{rgb}{1.0, 0.15, 0.99}
\definecolor{xgreen}{rgb}{0.8, 0.98, 0.8}
\definecolor{xorange}{rgb}{1.0, 0.733, 0.337}
\definecolor{xcyan}{rgb}{0.54, 0.98,1.0}
\definecolor{lightgrey}{rgb}{0.92, 0.96, 1.0}  
\definecolor{mypink}{rgb}{1, 0.9, 1} 
\definecolor{myblue}{rgb}{0.686, 0.796, 1.0} 
\definecolor{myyellow}{rgb}{0.9647, 1.0, 0.6157}
\definecolor{t1}{rgb}{1.0, 0.6039, 0.9961}
\definecolor{t2}{rgb}{0.8343, 0.5498, 0.9986}
\definecolor{t3}{rgb}{0.4902, 1.0, 0.4902}

\DeclareRobustCommand{\hlorange}[1]{{\sethlcolor{xorange}\hl{#1}}}

\DeclareRobustCommand{\hlblue}[1]{{\sethlcolor{myblue}\hl{#1}}}
\DeclareRobustCommand{\hlyellow}[1]{{\sethlcolor{myyellow}\hl{#1}}}
\DeclareRobustCommand{\hlone}[1]{{\sethlcolor{t1}\hl{#1}}}
\DeclareRobustCommand{\hltwo}[1]{{\sethlcolor{t2}\hl{#1}}}
\DeclareRobustCommand{\hlthree}[1]{{\sethlcolor{t3}\hl{#1}}}

\title{RUMAA: Repeat-Aware Unified Music Audio Analysis for Score-Performance Alignment, Transcription, and Mistake Detection}

\name{Sungkyun Chang,
      Simon Dixon,
      Emmanouil Benetos}
\address{Centre for Digital Music, Queen Mary University of London, UK
}

\begin{document}

\maketitle

\begin{abstract}
This study introduces \texttt{RUMAA}, a transformer-based framework for music performance analysis that unifies score-to-performance alignment, score-informed transcription, and mistake detection in a near end-to-end manner. Unlike prior methods addressing these tasks separately, \texttt{RUMAA} integrates them using pre-trained score and audio encoders and a novel tri-stream decoder capturing task interdependencies through proxy tasks. It aligns human-readable MusicXML scores with repeat symbols to full-length performance audio, overcoming traditional MIDI-based methods that rely on manually unfolded score-MIDI data with pre-specified repeat structures. \texttt{RUMAA} matches state-of-the-art alignment methods on non-repeated scores and outperforms them on scores with repeats in a public piano music dataset, while also delivering promising transcription and mistake detection results.
\end{abstract}

\begin{figure}[ht]
\centering
\includegraphics[width=1.0\columnwidth]{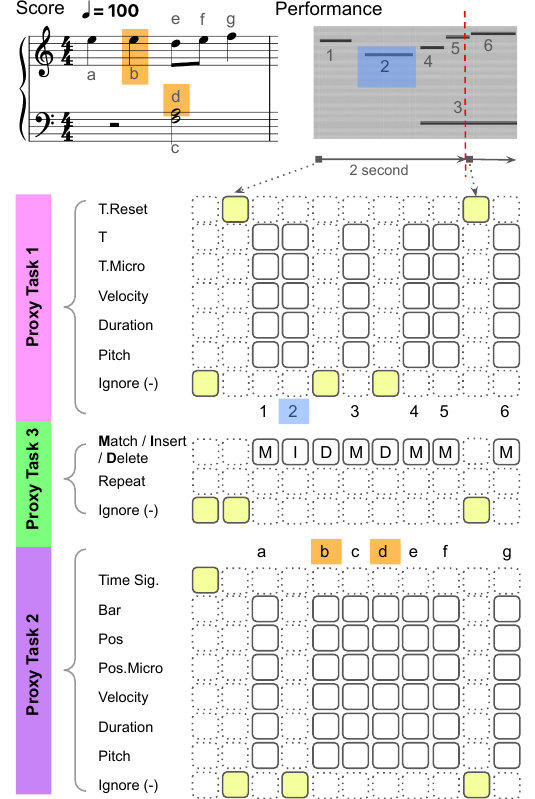}
\caption{\textbf{Token streams for proxy tasks.} \textbf{Top:} Example score and audio performance (visualized as a piano-roll), with orange highlights (\hlorange{missing}) for missing score notes and pastel blue highlights (\hlblue{extra}) for extra performance notes. \textbf{Bottom:} \texttt{RUMAA} decoder outputs for performance (Proxy Task 1), score (Proxy Task 2), and alignment (Proxy Task 3), as detailed in \cref{sec:tasks} and \cref{tab:token_def}. Yellow boxes (\hlyellow{$\Box$}) denote exclusive tokens, and dotted boxes indicate silence tokens. Patterns ensure grayscale readability. Best viewed in color.}
\label{fig:output_tokens}
\end{figure}

\section{Introduction}\label{sec:introduction}
Music Performance Analysis~\cite{lerch2019music} (MPA), a key area of Music Information Retrieval (MIR), investigates the relationship between a musical score and its performance across various genres and instruments. This study focuses on classical piano music, examining how performers adhere to the score, what deviations occur, and how audio translates into symbols. In classical music, where score fidelity is relatively critical, these questions support applications like tutoring~\cite{Zhang2024HowDataset} and assessment.

We address three essential tasks: \textit{score-to-performance alignment}, \textit{score-informed transcription}, and \textit{mistake detection}. Scores, with symbolic repeats and expressions, contrast with audio performances. Prior work treated these tasks separately—e.g., alignment mapping notes, transcription converting audio, and mistake detection identifying errors~\cite{nakamura2017performance,benetos2012score}—yet their interdependence is clear: alignment reveals matched or missing notes signaling mistakes, while transcription leverages alignment for accurate audio-to-symbol conversion. 

This interdependence suggests a unified approach would be more effective. Additionally, conventional MIDI-based representations cannot handle repeat symbols, which are often ignored in practice but strictly followed in formal performances. This poses challenges for MIDI-based alignment methods requiring manually verified, repeat-unfolded scores. In contrast, \texttt{RUMAA} directly processes repeat symbols.

\begin{figure*}[ht]
  \centering
  \includegraphics[width=0.97\textwidth]{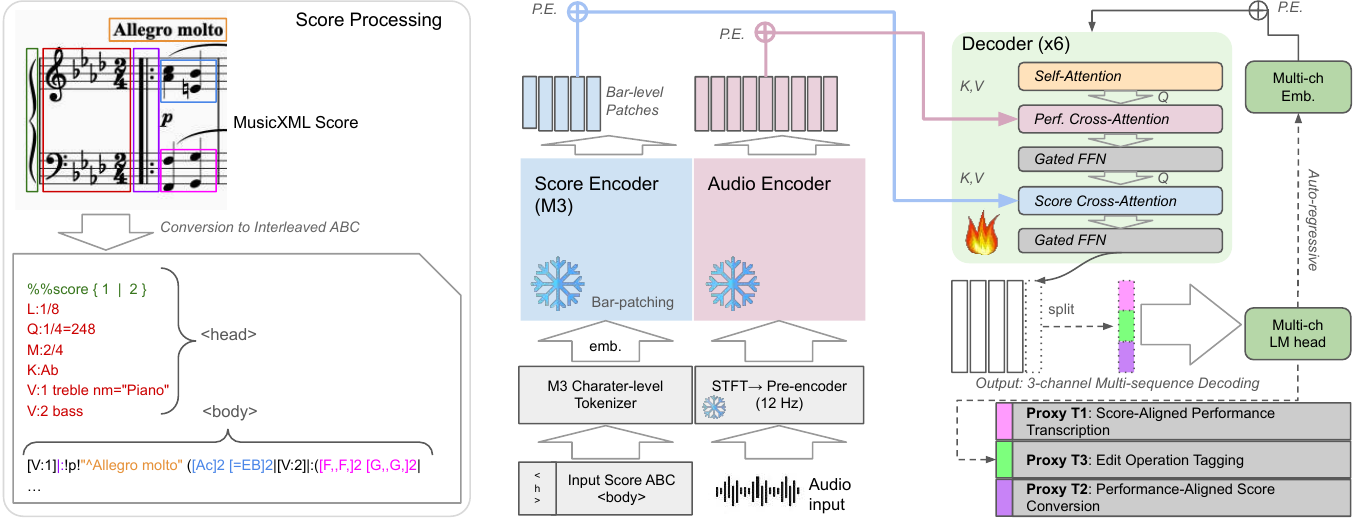}
  \caption{\textbf{Overview of \texttt{RUMAA} framework}. \textbf{Left}: The input MusicXML score is converted into an interleaved ABC representation (\cref{subsec:score_enc}). \textbf{Center}: The score and performance audio are encoded into score (\cref{subsec:score_enc}) and audio embeddings (\cref{subsec:perf_enc}). \textbf{Right}: The decoder (\cref{subsec:decoder}), conditioned on score and audio context, autoregressively generates multi-channel tokens. Each channel corresponds to a proxy subtask (\cref{sec:tasks}) related to our main tasks.}
  \label{fig:model_overview}
\end{figure*}

We propose \texttt{RUMAA} \textipa{[ru:ma:]} illustrated in \cref{fig:model_overview}, a transformer-based framework unifying these tasks with key features:
\begin{itemize}
\item \textit{Proxy-driven multi-task model}: Employs three proxy tasks to encode target task interdependencies, enhancing multi-task learning for transcription, alignment, and mistake detection.
\item \textit{Multimodal and crossmodal seq-to-seq}: Utilizes pre-trained score and audio encoders to jointly process symbolic scores and audio, enabling crossmodal reasoning and producing structured outputs that reflect task dependencies.
\item \textit{Repeat handling}: Handles repeats by aligning to performed structure without pre-unfolded scores, enabling adaptation to real-world performance variations.
\end{itemize}
This unified approach enables robust and flexible music performance analysis, bridging symbolic and audio modalities through proxy-driven, crossmodal learning.

\section{Related Work}\label{sec:related_work}
\textit{Score-performance alignment}~\cite[pp.~115--166]{muller2015fundamentals} employs cross-modal methods that directly align heterogeneous representations or symbolic methods that convert both modalities into a common space. Most methods adopt symbolic approaches, while cross-modal methods~\cite{muller2018cross, agrawal2021structure} align in the audio domain using features like \textit{chromagrams}, making note-level alignment challenging. Conventional DTW~\cite{keogh2005exact, simonetta2021audio, nasap-dataset} and HMM-based~\cite{nakamura2017performance} methods typically align transcribed symbolic sequences, incurring complexity and cascading errors from transcription.

\textit{Repeat handling} is a known issue~\cite{fremerey2010handling} in alignment methods. MIDI-based approaches struggle since MIDI lacks explicit repeat notations. JumpDTW~\cite{fremerey2010handling, Shan2020Improved} uniquely handles repeats by aligning image-domain scores with audio. Traditional methods rely on global structure analysis followed by local matching, while Transformer architectures may learn hierarchical patterns through multi-head attention, though long-context handling~\cite{dai2019transformer} remains challenging.

Recent deep learning methods include the GlueNote Transformer~\cite{peter24thegluenote}, which enhances DTW for symbolic alignment but requires external transcription. Chou et al.~\cite{chou_detecting_2025} proposed a Transformer for mistake detection, but we exclude it from baselines as it assumes pre-aligned 2-second score-audio pairs, making it unsuitable for inference on unaligned data.

\textit{Score-informed transcription} methods~\cite{benetos2012score, wang2017identifying} demonstrate that joint modeling enables accurate transcription and simultaneous alignment with mistake detection, inspiring our unified transformer-based approach.

\begin{table}[ht]
\centering
\caption{\textbf{Token definition.} Tokens marked with $\bigstar$ are exclusive, silencing other multi-class tokens in their channel. \texttt{<BOS>} and \texttt{<EOS>} are global tokens that silence all channels. $N$ includes a silence token per class.}
\label{tab:token_def}
\begin{tabular}{l c c l}
\toprule
\textbf{Token} & \textbf{Range} & \textbf{N} & \textbf{Description} \\
\midrule
\multicolumn{4}{l}{\textit{(Performance Tokens)}} \\
T (onset time) & 0:32 & 34 & 62.5 ms grid timing \\
T.Reset $\bigstar$ &  None & 2 & Reset T every 2 seconds \\
T.Micro & -5:5 & 11 & 6.25 ms adjustment  \\
Velocity & 1:32 & 32 & MIDI velocity, 4-unit steps \\
Duration & 0:48 & 49 & Up to 4s: 31.25/62.5/125 ms \\
Pitch & 21:109 & 89 & Piano MIDI pitch \\
- $\bigstar$ & None & 2 & Skip an extra note \\
\midrule
\multicolumn{4}{l}{\textit{(Score Tokens)}} \\
Time Sig. $\bigstar$ & 1/4 to 12/8 & 12 & Time signatures (top/bottom) \\
Bar & 0:50 & 51& Indexing bar \\
Pos & 0:32 & 33 & 40-tick (32nd note) grid \\
Pos.Micro & -5:5 & 11& 4-tick adjustment \\
Duration & 0:48 & 49 & 40/80/160-tick steps \\
Pitch & 21:109 & 89 & Piano MIDI pitch \\
- $\bigstar$ & None & 2 & Skip a missing note \\
\midrule
\multicolumn{4}{l}{\textit{(Alignment Tokens)}} \\
Insert/Delete/Match & None & 4 & Extra/Missing/Matched note \\
Repeat & None & 2 & Repeated notes or bar\\
- $\bigstar$ & None & 2 & Skip \\
\midrule
BOS, EOS & None & 2 & Begin and end of sequence \\

\bottomrule
\end{tabular}
\end{table}
\section{Proxy Task Design}\label{sec:tasks}
We address three interdependent tasks: score-performance alignment, score-informed transcription, and mistake detection. To efficiently learn these task relationships and enable their joint inference, we formulate a proxy task that learns to generate a tokenized alignment sequence of performance (\textbf{T1}) and score (\textbf{T2}) events, including explicit edit operations (\textbf{T3}) through our tri-stream decoding.

The core idea (Figure~\ref{fig:output_tokens}) is to create a strict one-to-one alignment by explicitly modeling extra or missing notes. This sequence integrates edit operations such as \texttt{<Match>}, \texttt{<Insert>}, and \texttt{<Delete>}. When a note is present in the performance but absent in the score (an insertion) or vice versa (a deletion), the exclusive token \texttt{<->} acts as a placeholder in the corresponding channel for the missing note. This maintains a consistent, perfectly aligned structure. 

Figure~\ref{fig:output_tokens} illustrates how each decoder channel corresponds to a proxy task. These tasks include: 
\begin{description}
  \item[\textbf{{T1.}}] \textbf{\hlone{Score-Aligned Performance Transcription}} outputs performance tokens with timing and note information.
  \item[\textbf{{T2.}}] \textbf{\hltwo{Performance-Aligned Score Conversion}} outputs score tokens following performance order.
  \item[\textbf{{T3.}}] \textbf{\hlthree{Edit Operation Tagging}} outputs alignment operation tokens.
\end{description}

This parallel decoding paradigm mirrors post-processed outputs from HMM/DTW-based methods using the Matchnote format~\cite{nakamura2017performance, nasap-dataset} adapted into a tokenized neural-friendly representation. While traditional approaches rely on global structure- or cluster-level alignment followed by local matching, our Transformer decoder directly learns these hierarchical patterns via cross-attention. Conditioning on full score input with barlines and repeats allows it to handle omissions and non-linear repeats without explicit rules.

Table~\ref{tab:token_def} provides the complete token definitions. Our tri-stream tokenization (Table~\ref{tab:token_def}) extends CP-Words~\cite{hsiao2021compound} by introducing structured tokens across three aligned streams: performance, score, and edit operations. Each note is encoded with onset, duration, and pitch tokens, including exclusive and silence tokens. Compared to serialized MIDI-like~\cite{mt3, bhandari2025improvnet} tokens, our design yields over 3× shorter sequences and enables simpler note-level alignment. By adopting two-level quantization—\texttt{<T>} and \texttt{<T.Micro>}\cite{lenz2024pertok}—and a timing reset token\cite{bradshaw2025scaling}, we achieve higher temporal resolution with fewer tokens than prior schemes, which typically require over 100 timing tokens per second. In contrast, our timing and duration representations are length-invariant, each using a fixed vocabulary of fewer than 50 tokens.

\section{Model}\label{sec:model}
\cref{fig:model_overview} illustrates the structure of \texttt{RUMAA}, which integrates pre-trained score and audio encoders with a custom decoder. Our design specifically addresses the training memory bottleneck of long audio inputs—unlike symbolic alignment models~\cite{peter24thegluenote} or short-audio transcription models (e.g., MT3~\cite{mt3, chang2024yourmt3+}) without score. We adopt: (1) efficient score representation using ABC notation with bar-level patching, (2) optimized audio encoding to reduce token length, and (3) hierarchical cross-attention decoder with tri-stream output strategy. The following subsections detail these components.

\subsection{Score Encoder}\label{subsec:score_enc}
As shown in the left panel of \cref{fig:model_overview}, the score encoder converts MusicXML~\cite{musicxml40} to ABC notation~\cite{walshaw2014abc}, representing each bar with up to 64 characters~\cite{wu2024clamp}. Both notes and key musical elements (dynamics, repeats, keys, time signatures, and pedal) are retained.

We adopt the pre-trained \texttt{M3} encoder from \texttt{CLaMP2}\cite{wu2024clamp}, which relies on character-level tokenization and bar-level patching to process ABC notation. Each character is encoded as a 12-dim embedding, and 64 embeddings per bar are stacked into a single 768-dim bar-level token. A 12-block Transformer with multi-head self-attention, pre-trained to capture musical events without losing note-level details from long sequences, produces a 768-dim representation\cite{wu2024clamp}. This output is linearly projected from 768 to 1,024 dimensions before being fed into the decoder described in Section~\ref{subsec:decoder}.

\subsection{Performance Audio Encoder}\label{subsec:perf_enc}
We use a spectrogram from 16 kHz mono audio (STFT: 2,048 samples window, 10ms hop). A pre-encoding layer (three ResNet blocks) processes features to produce 1024-dim outputs at 12 frames per second. Inspired by Music2Latent~\cite{pasini2024music2latent}, this lower frame rate efficiently supports longer sequences without degrading transcription performance, as verified in preliminary experiments.

The audio encoder is a 12-layer self-attention Transformer, pre-trained following YourMT3+~\cite{chang2024yourmt3+} with three modifications: adapted for lower-frame-rate inputs, re-implemented with flash attention, and pre-trained decoder predicting MIDI velocity tokens (0-127). In \texttt{RUMAA}, we reuse only the pre-trained and frozen audio encoder, which produces 1,024-dim representations.

\subsection{Decoder}\label{subsec:decoder}
The decoder, shown as the green box in the center of \cref{fig:model_overview}, is a 6-block Transformer autoregressively generating tri-stream outputs conditioned on audio (\cref{subsec:perf_enc}) and score (\cref{subsec:score_enc}) features. 

Our key architectural choice is hierarchical cross-attention, which involves separate conditioning for audio then score, in contrast to standard concatenation~\cite{chou_detecting_2025} or prepending input. Each block processes shifted outputs via self-attention, then cross-attention with audio, then score. This sequential approach resembles iterative transcription/score-following and is more efficient than simple context concatenation. Replacing hierarchical attention with simple concatenation leads to ~1\% performance drop despite similar model size.

Built on TorchScale~\cite{torchscale}, the Transformer uses a GatedFFN~\cite{raffel2020exploring, ding2023longnet} and extends standard decoder blocks with an additional cross-attention layer as mentioned above. The final 1024-dime latent representation is divided into three streams (1024 // 3), each processed by a multi-sequence language model (LM) head~\cite{hsiao2021compound} to generate parallel sequences. Each head corresponds to a proxy task defined in Section~\ref{sec:tasks}: \textbf{T1}, \textbf{T2}, \textbf{T3}. The third head also has the potential to decode extended compound tokens, such as beat, tempo, or other attributes relevant to music performance modeling. Note that tokens with the same name across score and performance channels (e.g., \texttt{<Pitch>}; see \cref{tab:token_def}) use separate embeddings and do not share weights.

\section{Experimental Setup}\label{sec:experimental_setup}

\subsection{Data Preparation and Training}\label{subsec:dataset}

For pre-training the audio encoder, we use the YourMT3 dataset~\cite{chang2024yourmt3+}, combining 10 public multi-instrument transcription datasets including Maestro~\cite{maestro}, reserving the official test split for evaluation. While prioritizing piano, we maintain multi-instrument pre-training, as initial findings show that mapping speech to singing and non-musical or percussive sounds to drums—or ignoring them—boosts robustness in live piano transcription.

For post-training the decoder, we use the \texttt{\texttt{(n)ASAP}} dataset~\cite{nasap-dataset}, a \texttt{Maestro}\cite{maestro} subset with 222 MusicXML~\cite{musicxml40} scores and 519 piano performances, including MIDI, audio, and manually verified note alignments~\cite{nasap-dataset}. We held out 20 movements and 50 recordings from six composers—Bach, Beethoven, Chopin, Haydn, Liszt, and Schubert—as the test set. MusicXML scores are converted to ABC-interleaved format~\cite{wu2024clamp} for the score encoder. Tri-stream tokens for the proxy tasks derived from \texttt{\texttt{(n)ASAP}}’s alignment, MIDI, and score data. We randomly sample one-minute audio segments and extract up to 50 bars of score content that cover the corresponding passage.

To augment \texttt{\texttt{(n)ASAP}}’s semi-professional performances, which limit mistake learning and rarely include repeats in 1-minute segments, we create five score-modulated versions, altering 10\% of notes via pitch modulation (±5 semitons) or deletion, and five performance-modulated versions using Piano-SynMist~\cite{morsi2024simulating} and MIDI-DDSP~\cite{wu2021midi}. This expands the dataset tenfold for robust proxy task training. For repeat simulation, we add repeat symbols to random bars in 20\% of ABC-interleaved scores lacking repeats, repeating the audio. 

To evaluate score-performance alignment, we use the revised \texttt{Vienna}~\cite{vienna4x22_rematched}, with high-quality piano performances, transcriptions, MusicXML scores, and manual alignments. Score-informed transcription is tested on \texttt{\texttt{(n)ASAP}}~\cite{nasap-dataset}, score-free transcription on \texttt{Maestro}\cite{maestro}. Lastly, \texttt{STPD}~\cite{benetos2012score} is used for mistake detection benchmark, with its score-MIDI converted to the ABC-interleaved format. All the evaluation datasets are isolated from the training dataset.

We trained our decoder alongside an off-the-shelf M3~\cite{wu2024clamp} score encoder and a pre-trained performance audio encoder described in \cref{subsec:score_enc}.  For this post-training, we adopted AdamW-Scale optimizer~\cite{adamwscale} and scheduler from the prior work~\cite{chang2024yourmt3+}. Training takes on three A6000 GPUs or H100 GPUs using a cosine schedule with initial and final learning rates of [1e-02, 1e-05] and a 1,000-step warm-up from 1e-03, spanning approximately two days.

\subsection{Evaluation Metrics}\label{subsec:metrics}
Since our task includes transcription, we adopt a widely used ±50 ms onset tolerance~\cite{raffel2014mir_eval} across all evaluations.

For score-to-performance alignment, we employ the note-level \textbf{F\textsubscript{align}} metric~\cite{nasap-dataset}. This metric evaluates matched note pairs and inserted/deleted notes as \textit{True Positives}, unmatched predicted notes as \textit{False Positives}, and missing ground-truth notes as \textit{False Negatives}. Originally designed for symbolic alignment tasks without repetitions, we adapt it by redefining repeated notes to be counted independently.

For score-informed transcription, we utilize two metrics derived from \texttt{mir\_eval}~\cite{raffel2014mir_eval}: \textbf{F\textsubscript{on}} to evaluate note onset detection and \textbf{F\textsubscript{off-vel}} to assess combined onset, offset, and velocity detection. Additionally, \textbf{MAE\textsubscript{vel}} measures the mean absolute error of velocity (0--127).

For the mistake detection task in \cref{tab:result_benetos_mistake_detection}, we adopt four metrics consistent with prior studies~\cite{wang2017identifying, ewert2016score}: \textbf{F\textsubscript{correct}}, \textbf{Acc\textsubscript{correct}} for correctly played notes, \textbf{F\textsubscript{extra}}, \textbf{Acc\textsubscript{extra}} for erroneous notes not in the score, and \textbf{F\textsubscript{missed}}, \textbf{Acc\textsubscript{missed}} for notes omitted from the performance despite being in the score.

\section{Result}\label{sec:result}

\begin{table}[ht]
\centering
\caption{\textbf{Note-level alignment (F\textsubscript{align}) on the piano score-to-performance task.} ``w/o repeat'' indicates evaluation with all songs from \texttt{Vienna}~\cite{vienna4x22_rematched} dataset, while ``w/ repeat'' indicates evaluation with two songs containing repeat symbols (\textit{Mozart K331}, \textit{Schubert D783}).}
\label{tab:result_alignment}
\begin{tabular}{lcc}
\toprule
\textbf{Model} &\textbf{w/o repeat}  & \textbf{w/ repeat} \\ 
\midrule
\small\textit{(symbolic alignment)} & & \\
Nakamura HMM~\cite{nakamura2017performance}& \textbf{99.0} & \underline{36.4} \\
hDTW+sym~\cite{nasap-dataset} & 98.5 & 28.2 \\
GlueNote Transformer~\cite{peter24thegluenote} & 98.5 & 12.7 \\ 
\midrule
\small\textit{(symbolic-audio alignment)} & & \\
AMT~\cite{chang2024yourmt3+} + Nakamura~\cite{nakamura2017performance}  & 97.4 & 31.8\\
AMT~\cite{chang2024yourmt3+} + hDTW+sym~\cite{nasap-dataset} & 96.9& 26.5\\
AMT~\cite{chang2024yourmt3+} + GlueNote~\cite{peter24thegluenote} & 96.9 & 26.3 \\
RUMAA (ours) & \underline{98.4} & \textbf{98.4} \\ 
\bottomrule
\end{tabular}
\end{table}

This section reports the evaluation of \texttt{RUMAA} on score-to-performance alignment, score-informed transcription, and mistake detection. Results are summarized in \cref{tab:result_alignment}–\ref{tab:result_benetos_mistake_detection}, showing its performance across these tasks within a unified framework, with particular focus on scores with repeats and score-informed processing.

\subsection{Score-to-Performance Alignment}
For the score-to-performance alignment task, \cref{tab:result_alignment} reports F\textsubscript{align} values on the \texttt{Vienna}~\cite{vienna4x22_rematched} piano dataset. The ``without repeat'' setting uses unfolded scores across all songs, while the ``with repeat'' setting applies original scores to songs with repeat symbols. The upper section, symbolic alignment, includes methods HMM~\cite{nakamura2017performance}, hDTW+sym~\cite{nasap-dataset}, and GlueNote Transformer~\cite{peter24thegluenote}, aligning ground truth performance MIDI with unfolded score MIDI under ideal conditions without transcription errors nor repeats. Nakamura HMM achieves the highest performance on non-repeated scores, with \texttt{RUMAA} trailing by less than 1\% (F1-score 98.4). On scores with repeats, however, these methods drop by up to 87\%, while \texttt{RUMAA} maintains 98.4. Conventional methods~\cite{nakamura2017performance, nasap-dataset, peter24thegluenote} using score-MIDI cannot interpret repeat symbols, often aligning only the beginning and end of repeated scores. This is particularly noticeable in GlueNote, which struggles more with repeats.

The lower section, symbolic-audio alignment, reflects alignment errors that inherently include prior transcription errors, as the alignment is performed on transcribed symbolic data. On repeated scores, these methods drop by up to 87\%, while \texttt{RUMAA} maintains 98.4. In symbolic-audio alignment, with transcription errors, baselines combining an external automatic  music transcription (AMT) ~\cite{chang2024yourmt3+} with prior methods decline by up to 70\% on repeats, but \texttt{RUMAA} outperforms them by over threefold, enhancing practicality without an external transcriber.

\begin{table}[ht]
    \centering
    \caption{\textbf{Note-level transcription performance (Onset, Offset-Velocity F1 and Velocity MAE) on the score-informed piano transcription task}. ``w/o score'' indicates evaluation without score guidance, while ``Score-informed'' uses aligned score data from the \texttt{\texttt{(n)ASAP}} dataset~\cite{nasap-dataset} on the same test set. Models marked with an asterisk (*) are trained for multi-instrument transcription, while the others are piano-only.}
    \label{tab:result_informed_transcription}
    \begin{tabular}{lccc}
        \toprule
        \textbf{Model} & \textbf{F\textsubscript{on} $\uparrow$} & \textbf{F\textsubscript{off-vel} $\uparrow$} & \textbf{MAE\textsubscript{vel} $\downarrow$} \\
        \midrule
        \small\textit{(w/o score: Maestro)} & & & \\
        hFT-T~\cite{toyama2023} & 97.4 & 89.5 & - \\
        IS-CRF~\cite{yan2024scoring} & \underline{98.3} & \underline{93.0} & - \\
        MT3$^{*}$~\cite{mt3}  & 84.9 & - & - \\
        YourMT3+$^{*}$~\cite{chang2024yourmt3+} & 97.0 & -  & - \\
        RUMAA$^{*}$  & 96.1 & 76.0 & 3.5\\
        \midrule
        \midrule
        \small\textit{(w/o score: (n)ASAP)} & & & \\
        RUMAA$^{*}$  & 95.9 & 75.8 & 4.6\\
        \midrule
        \small\textit{(score-informed: (n)ASAP)} & & & \\
        RUMAA & \textbf{99.1} & \textbf{93.6} & \textbf{4.0} \\
        \bottomrule
    \end{tabular}

\end{table}
\begin{table}[ht]
    \centering
    \caption{\textbf{Benchmark for piano mistake detection on the \texttt{STPD}~\cite{benetos2012score} dataset.}}
    \label{tab:result_benetos_mistake_detection}
    \sisetup{
        reset-text-series = false, 
        text-series-to-math = true, 
        mode=text,
        tight-spacing=true,
        round-mode=places,
        round-precision=2,
        table-format=2.2,
        table-number-alignment=center,
    }
    \setlength{\tabcolsep}{3.5pt}
    \begin{tabular}{lccccccc}
        \toprule
        \textbf{Model} & \textbf{w/o Score} & \multicolumn{6}{c}{\textbf{Score-informed}} \\
        \cmidrule(lr){2-2} \cmidrule(lr){3-8}
        & F\textsubscript{on} & F\textsubscript{correct} & F\textsubscript{extra} & F\textsubscript{missed} & A\textsubscript{correct} & A\textsubscript{extra} & A\textsubscript{missed} \\
        \midrule
        Wang~\cite{wang2017identifying} & -- & 99.2 & \underline{84.9} & 92.6 & 98.4 & \underline{75.2} & 86.9 \\
        Ewert~\cite{ewert2016score} & -- & \underline{99.3} & 77.0 & \underline{94.5} & \underline{98.6} & 64.0 & \underline{89.9} \\
        Benetos~\cite{benetos2012score} & 91.1 & -- & -- & -- & 93.2 & 60.5 & 49.2 \\
        RUMAA & \textbf{92.8} & \textbf{99.5} & \textbf{89.2} & \textbf{95.3} & \textbf{98.7} & \textbf{80.3} & \textbf{90.4} \\
        \bottomrule
    \end{tabular}
\end{table}

\subsection{Score-informed Piano Transcription}
\cref{tab:result_informed_transcription} evaluates score-informed transcription on the \texttt{(n)ASAP} dataset~\cite{nasap-dataset}, reporting onset F1, offset-velocity F1, and velocity MAE. Without score guidance on \texttt{Maestro}, \texttt{RUMAA} competes with top models like hFT-T~\cite{toyama2023} and YourMT3+\cite{chang2024yourmt3+}, though it falls short of IS-CRF\cite{yan2024scoring} by about 2\%. \texttt{RUMAA}'s note offset-velocity prediction performance drops by up to 18\% compared to other models. This may stem from inconsistent offset annotations in the multi-instrument dataset used to pre-train our audio encoder, unlike competing models optimized for \texttt{Maestro}. However, with score information on \texttt{(n)ASAP}, the proposed \texttt{RUMAA} achieves a near-perfect onset F1 score (99.1), surpassing all baselines by a clear margin, demonstrating that leveraging score information significantly enhances transcription performance.



\subsection{Piano Mistake Detection}
For mistake detection, \cref{tab:result_benetos_mistake_detection} presents a performance comparison on the \texttt{STPD}\cite{benetos2012score} piano dataset alongside previous NMF-based methods. \texttt{RUMAA} outperforms prior NMF-based approaches\cite{wang2017identifying, ewert2016score, benetos2012score} in mistake detection. Without score guidance, it surpasses Benetos et al.~\cite{benetos2012score} by approximately 2\%; with score information, it achieves up to a 15\% improvement in detecting extra and missed notes while maintaining top accuracy metrics.


\section{Discussion and Future Work}\label{sec:limitation}
\texttt{RUMAA} demonstrates that a unified model for score-performance alignment, transcription, and mistake detection achieves strong performance across all tasks. Notably, it performs well on scores with repeats, indicating effective modeling of MusicXML repeat structures. A key advantage of the unified audio and score encoders is reduced transcription error via score-informed decoding, in contrast to conventional AMT + HMM/DTW pipelines.

However, the model struggles with long audio sequences (over one minute) due to cross-chunk memory limits, restricting its use on extended real-world recordings~\cite{flossmann2010magaloff} for alignment and mistake detection. Evaluations were also limited to relatively clean, single-instrument data, and online processing remains unexplored.

To address these limitations, future work could proceed along one of two avenues: developing memory-augmented architectures~\cite{graves2014neural} to overcome sequence length constraints, or enhancing the model's generalizability by extending it to multi-instrument scores and more diverse, real-world recordings.



\clearpage
\section{Acknowledgment}
\label{sec:ack}
This research partially utilized AI Industrial Convergence Cluster supported by the Ministry of Science and ICT of Korea, and Gwangju Metropolitan
City.


\bibliographystyle{IEEEtran}
\bibliography{references}

\end{document}